\documentclass[12pt]{elsarticle}
\makeatletter
\def\ps@pprintTitle{%
 \let\@oddhead\@empty
 \let\@evenhead\@empty
 \def\@oddfoot{\centerline{\thepage}}%
 \let\@evenfoot\@oddfoot}
\makeatother

\usepackage{amssymb}
\usepackage{algorithm}
\usepackage{algorithmic}
\usepackage{amsmath}
\usepackage{amssymb}
\usepackage{array}
\usepackage{graphicx}
\usepackage{subfigure}
\usepackage{wrapfig}
\usepackage{rotating}
\usepackage{longtable}
\usepackage{amsmath,epsfig,times}

\usepackage{amsmath}
\usepackage{amssymb}

\usepackage{psfrag}
\usepackage{graphicx}

\usepackage{multirow}
\usepackage{array}
\usepackage{enumerate}

\usepackage{color}
\usepackage{caption}
\usepackage{ltxtable, tabularx, longtable}
\usepackage{amssymb}
\usepackage{atbegshi}
\usepackage{mathtools}
\usepackage{float}

\begin{document}

\begin{frontmatter}

\title{Amateur Drones Detection: A machine learning approach utilizing the acoustic signals in the presence of strong interference}

\author{Zahoor Uddin\fnref{myfootnote} }
\fntext[myfootnote1]{COMSATS University Islamabad, Wah Campus, Pakistan; zahooruddin79@gmail.com, mohammadaltaf@ciitwah.edu.pk}

\author{ Muhammad Altaf \fnref{myfootnote} }

\author{ Muhammad Bilal \fnref{myfootnote2}\corref{mycorrespondingauthor} }
\fntext[myfootnote2]{ Div. of Computer and Electronics Systems Engineering,Hankuk University of Foreign Studies, Yongin-si, 17035, South Korea; m.bilal@ieee.org}

\author{  Lewis Nkenyereye \fnref{myfootnote3}\corref{mycorrespondingauthor} }
\fntext[myfootnote3]{Department of Computer and Information Security, Sejong University, Seoul, South Korea;  nkenyele@sejong.ac.kr}

\author{Ali Kashif Bashir\fnref{myfootnote4} }
\fntext[myfootnote4]{Department of Computing and Mathematics, Manchester Metropolitan University, United Kingdom; dr.alikashif.b@ieee.org\\
\\
\colorbox{green}{Please cite this article as: Z. Uddin, M. Altaf, M. Bilal et al.,Amateur Drones Detection: A machine} \\ \colorbox{green}{ learning approach utilizing the acoustic signals in the presence of strong interference, Computer} \\ \colorbox{green}{  Communications (2020), https://doi.org/10.1016/j.comcom.2020.02.065.}}

\cortext[mycorrespondingauthor]{Corresponding authors}
\cortext[myfootnote1]{This work was supported by in part by HUFS research funds 2020 and in part by the Sejong University New Faculty Program through the National Research Foundation of Korea NRF (No. 20180372: Privacy Preserving and Authentication Techniques for Service Oriented Applications in 5G Fog-based IoV)}

\begin{abstract}
Owing to small size, sensing capabilities and autonomous nature, the Unmanned Air Vehicles (UAVs) have enormous applications in various areas e.g., remote sensing, navigation, archaeology, journalism, environmental science, and agriculture. However, the un-monitored deployment of UAVs called the amateur drones (AmDr) can lead to serious security threats and risk to human life and infrastructure. Therefore, timely detection of the AmDr is essential for the protection and security of sensitive organizations, human life and other vital infrastructure.  AmDrs can be detected using different techniques based on sound, video, thermal, and radio frequencies. However, the performance of these techniques is limited in sever atmospheric conditions. In this paper, we propose an efficient un-supervise machine learning approach of independent component analysis (ICA) to detect various acoustic signals i.e., sounds of bird, airplanes, thunderstorm, rain, wind and the UAVs in practical scenario. After unmixing the signals, the features like Mel Frequency Cepstral Coefficients (MFCC), the power spectral density (PSD) and the Root Mean Square Value (RMS) of the PSD are extracted by using ICA. The PSD and the RMS of PSD signals are extracted by first passing the signals from octave band filter banks. Based on the above features the signals are classified using Support Vector Machines (SVM)and K Nearest Neighbor (KNN)to detect the presence or absence of AmDr. Unique feature of the proposed technique is the detection of a single or multiple AmDrs at a time in the presence of multiple acoustic interfering signals. The proposed technique is verified through extensive simulations and it is observed that the RMS values of PSD with KNN performs better than the MFCC with KNN and SVM.

\end{abstract}


\end{frontmatter}

\section{Introduction}
The modern days Unmanned Air vehicles (UAVs) also called drones are equipped with advanced electronics, control and communication technologies.  These small flying processing systems unite with communication capabilities has brought ensures advancement in various fields. The applications of UAvs mainly includes rescue and relief operations in post disaster situation, data demand management of wireless communication system during special events of large gatherings, remote sensing applications  and military applications etc.,  as summarized in Table \ref{tab:table1}.

\begin{table}[H]
\caption{Applications of UAVs in various areas.}
\centering
\begin{tabular}{ccc}
\hline\noalign{\smallskip}
\textbf{Ref. No.} & \textbf{Year of publication} & \textbf{Application area}\\ 
\noalign{\smallskip}\hline\noalign{\smallskip}
\cite{wc1} & 2016 & Disaster relief \\ 
\cite{wc2}& 2016 & Interference reduction in communication\\ 
\cite{wc3} & 2016 &Data rate enhancement in communication \\ 
\cite{wc4} & 2016 &Data demand management during various events\\ 
\cite{wc5} & 2017 & Wireless coverage management during disaster\\
\cite{wc6} & 2013 &Archaeological surveys \\ 
\cite{wc7} & 2013 &Soil erosion measurement \\ 
\cite{wc8} & 2013 &Ortho-photo-imaging \\ 
\cite{wc9} & 2015 & Agriculture\\ 
\cite{wc10} & 2015 & Archaeology\\ 
\cite{wc11} & 2016 & Environment scanner\\ 
\cite{wc12} & 2017 &Scene detection through image processing \\ 
\cite{wc13} & 2013 & Remote sensing applications\\ 
\cite{wc14} & 2013 & Military observation of a region to locate enemy \\ 
\cite{wc15} & 2013 & Earthquake\\ 
\cite{wc16} & 2017 & Intelligent navigation \\ 
\cite{wc17} & 2014 & Survey and classification of trees  \\ 
\cite{wc18} & 2014 & Journalism  \\ 
\noalign{\smallskip}\hline
\end{tabular}
\label{tab:table1}
\end{table}

Although, drones having countless applications in various areas, but its ungoverned deployment causes serious security problems \cite{com1} and require intelligent transportation techniques \cite{com2}. It can be used for transfer of unlawful material like explosives or can violate the boundary of security sensitive organizations. These illegalities might be floated  from illegal organizations inclusive of terrorist organizations. Drone detection is extensively performed and presented in the research literature. In \cite{ad1}, the authors proposed point to point architecture that can detect presence of a single  AmDr in the sensing field. However, due to its limited detection capability it is not recommended in highly sensitive  areas, as the presence of multiple drones goes undetected. In practical scenario, the existence of multiple drone in the restricted area is highly probable, the proposed scheme in \cite{ad1} employs the point to point detection mechanism, which is capable of detecting single drone, but fail to detect the existence of multiple drone.  Surveillance drone based AmDr detection is performed in \cite{ad2}, where the authors developed a technique based on strength of the radio frequency (RF) signals and modulation classification of the signals. In \cite{ad3}  cognitive internet of things is utilized for AmDr detection based on sound, video, thermal, and radio RF signals. Image processing based AmDr detection is performed in \cite{ad5,ad6}. Various machine learning technique are employed for AmDr detection in \cite{ad4,ad10}. The implementation of radar technology is discussed in \cite{ad7} for AmDr detection. In \cite{ad8}, an RF based experimental setup is developed for AmDr detection.  Radio access network based AmDr detection is performed in \cite{ad9}. In reference \cite{ad11} the authors presented sound correlation based drone detection technique. Video based detection is performed in \cite{ad12}.   Hidden Markov model is utilized for drone detection in \cite{ad13}. An anti-Drone system is developed in \cite{xshi} to detect a single drone in the surrounding field by utilizing audio, video and the RF signals. This work also utilized independently recorded signals without considering the interfering sources while utilizing the audio signals. However, all these techniques are mainly focusing on single drone detection, which limits the applicability of these schemes in practical applications where presence of multiple drones is highly likely. These techniques are summarized in Table \ref{tab:table2}.

\begin{figure}[H]
\centering
  \includegraphics[width=4in]{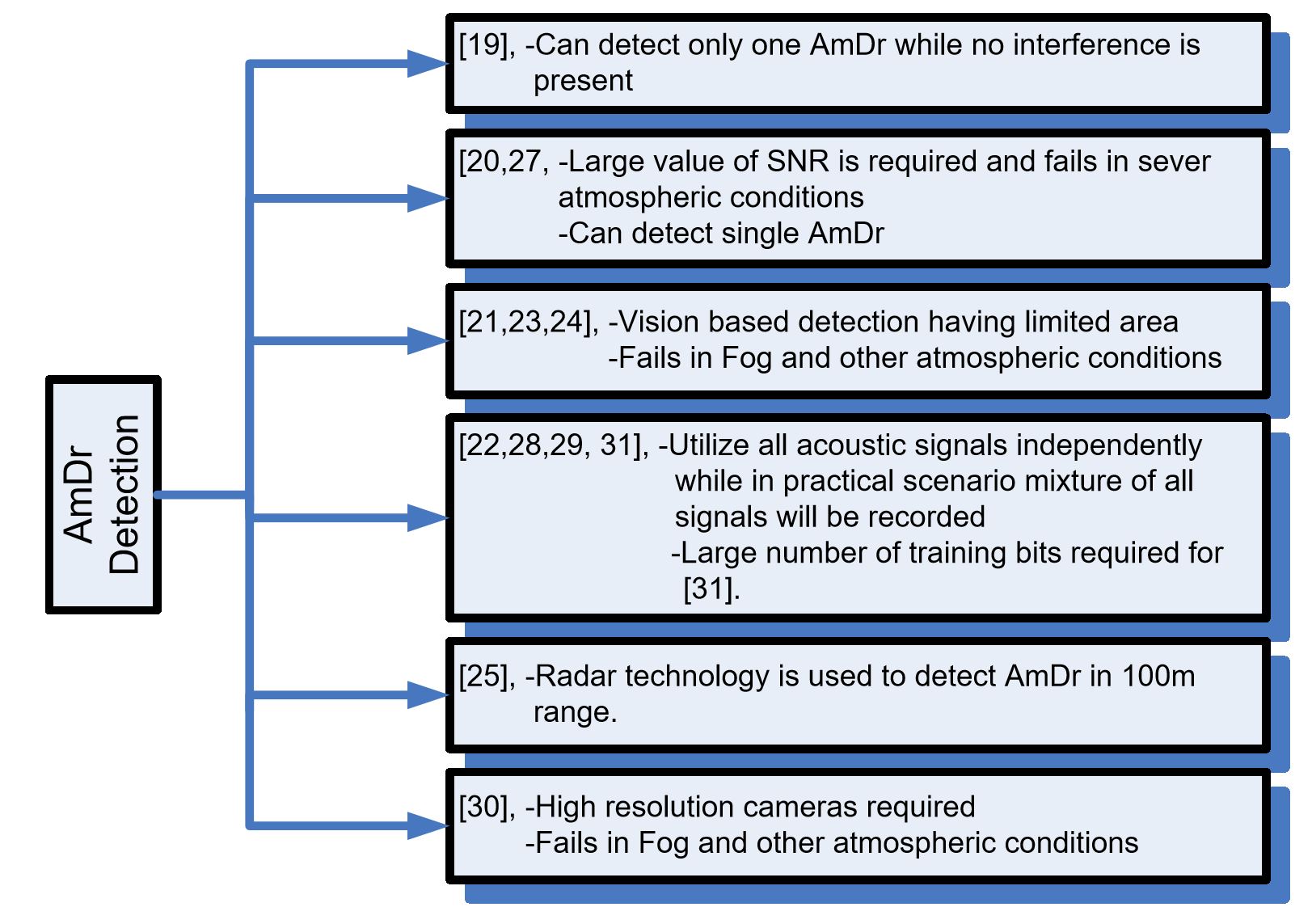}
 
  \caption{Drawbacks of the existing AmDr detection techniques}
    \label{Fig. 1}
\end{figure}

Furthermore, from the above discussion it can be noted that the RF communication, acoustic measurement, image and video signal processing techniques are employed for the AmDr detection. However, RF based detection fails in sever atmospheric conditions and has limitation of detecting smaller and variable shape drones. Similarly, the Image and video based techniques require high performance cameras and computationally efficient circuitry, hence is a very costly solution. Moreover, these image and video based technique has limitations due to its fixed orientation. The sound based detection is more practical, but various interfering sound sources like, birds, aeroplane, wind, thunderstorm etc., makes it more challenging. A recent research \cite{ad4} has proposed sound based AmDr detection, but has considered all the sounds independently. In practical scenario the recorded sounds are normally mixtures of all the existing sounds. Therefore, the work presented in \cite{ad4} is not applicable in practical scenario as the technique presented is for a single drone detection and that too in the presence of interfering sources. The shortcoming of the existing AmDr detection techniques is presented in Figure \ref{Fig. 1}.

\begin{table}[H]
\centering
\caption{Overview of the existing AmDr detection techniques.}
\begin{tabular}{ccccc}
\hline\noalign{\smallskip}
\textbf{Ref. No.} & \textbf{Year}  & \textbf{AmDr detetcion Technology}  & \textbf{Multiple}   \\
 &  &   &  \textbf{UVAs detection} \\ 
 \noalign{\smallskip}\hline\noalign{\smallskip}
\cite{ad1} & 2018 & Ad-hoc network architectures & No \\ 
\cite{ad2} & 2017 & RF signal strength &  No\\ 
\cite{ad3} & 2017 & Cognitive Internet of Things & No  \\ 
\cite{ad5} & 2018 & Image processing & No   \\ 
\cite{ad6} & 2017 & Image processing & No   \\ 
\cite{ad4} & 2018 & Machine learning &  No  \\ 
\cite{ad10} & 2017 & Machine Learning & No  \\ 
\cite{ad7} & 2016 & Radar technology & No  \\ 
\cite{ad8} & 2016 &RF based experimental system  & No  \\ 
\cite{ad9} & 2018 &Radio access network  & No \\ 
\cite{ad11} & 2016 & Sound correlation & No  \\ 
\cite{ad12} & 2017 & Video signal processing & No  \\ 
\cite{ad13} & 2018 & Hidden Markov model  & No  \\ 
\noalign{\smallskip}\hline
\end{tabular}
\label{tab:table2}
\end{table}

In this paper, a more practical approach of the AmDr detection is proposed. The proposed approach is a combination of the supervised and the un-supervised machine learning techniques. We utilize the  independent component analysis (ICA) \cite{c2,c3,c4} to efficiently detect the presence of multiple AmDrs in the sensing field. ICA is an un-supervised machine learning technique \cite{c5,c6}, used for separating the mixed recorded acoustic signals, which makes the ICA a good candidate for detection of single or multiple AmDrs in the presence of strong interfering sources. In precise, in our proposed scheme the ICA post processed signals are further processed for features extraction, which are then processed through the supervised machine learning techniques for classification to detect presence of the AmDrs. Further, the MFCC algorithms \cite{LMCC} are used to extract Mel Frequency Cepstral Coefficients from the separated, independent and non-Gaussian source signals that are further processed with SVM and the KNN \cite{LSVM} for classification into drone and non-drone signals. In a second approach the ICA post processed signals are passed through a series of band pass filters, tuned to octave band frequencies. The Power Spectral Density (PSD) and Root Mean Square (RMS) values of the source signals in each frequency band are calculated. These PSD and RMS values are fed as features vectors to SVM and KNN for identification of the AmDrs. In addition, this is the first time to propose the ICA based AmDr detection technique with filter banks in octave frequency bands. This implementation is feasible for various practical scenarios as it can detect both single and multiple AmDrs in the presence of various interfering signals e.g., aeroplanes, birds, wind, rain and thunderstorm sounds.

Rest of the paper is organized as : Section 2 contains the system model of the proposed work. Section \ref{sec:data} presents the data description. Section \ref{sec:prop} demonstrates the proposed technique, while  experimental verification is given in Section \ref{sec:Exp}. Finally, concluding remarks  are given in Section \ref{sec:conclude}.

\emph{Notations}: Lowercase normal letters represent scalars , lowercase boldface letters are used for vectors (e.g., ${\textbf{x}}$, ${\textbf{y}}$, ${\textbf{z}}$, ...), and uppercase boldface letters represent matrices(e.g., ${\textbf{X}}$, $\textbf{Y}$, $\textbf{Z}$,...).

\section{The System Model}
\begin{figure}[H]
 \centering

  \includegraphics[width=3in]{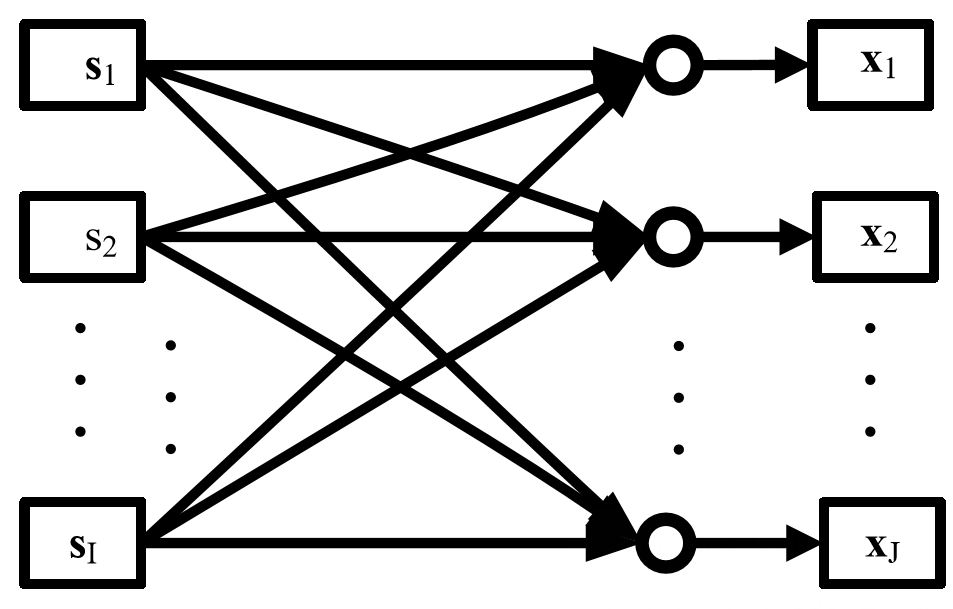}

  \caption{System model of the proposed work. This figure shows the recording mechanism of the AmDrs and the interfering source signals.}
    \label{Figure 2}
\end{figure}

The main goal of this research is to detect the presence of single or multiple AmDrs in the sensing field through utilizing the acoustic signals. The AmDr detection is carried out in the presence of multiple interfering sources like aeroplanes, birds, wind, rain and thunderstorm sounds. In practical scenario, the sensors record mixtures of the sounds present in the field. Let we have $I$ number of AmDrs and interfering sources that are recorded by $J$ number of microphones. For sake of simplifying the understanding let $I=J$. The audio source signals are $\textbf{s}_1, \textbf{s}_2,.....,\textbf{s}_I$, where $\textbf{s}_i=[s_{i1}, s_{i2}, .... ,s_{iL}]$, and the recorded mixed signals are $\textbf{x}_1, \textbf{x}_2,....,\textbf{x}_J$. The phenomenon of recording multiple mixed signals is shown in Figure \ref{Figure 2}. Mathematically, the mixed recorded data can be represented as
\begin{equation}\label{sinr}
\textbf{X}=\textbf{A}\textbf{S}
\end{equation}
where $\textbf{X}$ is $J \times L$ mixed data matrix, contains mixtures of the AmDrs and interfering sources, $L$ represents length of the processing data blocks, $\textbf{A}$ is $I \times J$ mixing matrix and $\textbf{S}$ is $I \times L$ source data matrix. The mixing matrix represents the mixing phenomenon and contains the mixing coefficients. The source data matrix contains the original sound signals of the AmDrs and the interfering sources. Hence, the proposed algorithm utilize the mixed data to detect the presence or absence of the AmDrs. The mixed data is first un-mixed through ICA algorithm \cite{d,e} and denoted by $\textbf{y}_1, \textbf{y}_2, ... ,\textbf{y}_J$ and then processed through the features extraction and classification techniques for the AmDrs detection.
There are some conditions utilized in this research for the AmDr detection, as given below

\begin{itemize}
  \item All source signals "$\textbf{s}_i$"  are assumed to be statistically independent. This condition is practically valid because all the source signals are generated by different sources.
  \item All "$\textbf{s}_i$" have non-Gaussian distributions, that is true in case of audio signals. This assumption is utilized because according to Central Limit Theorem, if we mix two or more non-Gaussian signals the resultant mixed signals will be more Gaussian and by maximizing the on-Gaussianity we will get the resultant un-mixed signals.

  \item "$\textbf{A}$" is assumed to be square for the sake of simplicity. This condition is easy to achieve. If the number of sensors become equal to the number of source signals, "$\textbf{A}$" becomes square. The square mixing matrix provides equal number of un-known variables and equations.
\end{itemize}

\section{The data description} \label{sec:data}
\begin{figure}[H]
 \centering

  \includegraphics[width=3.5in]{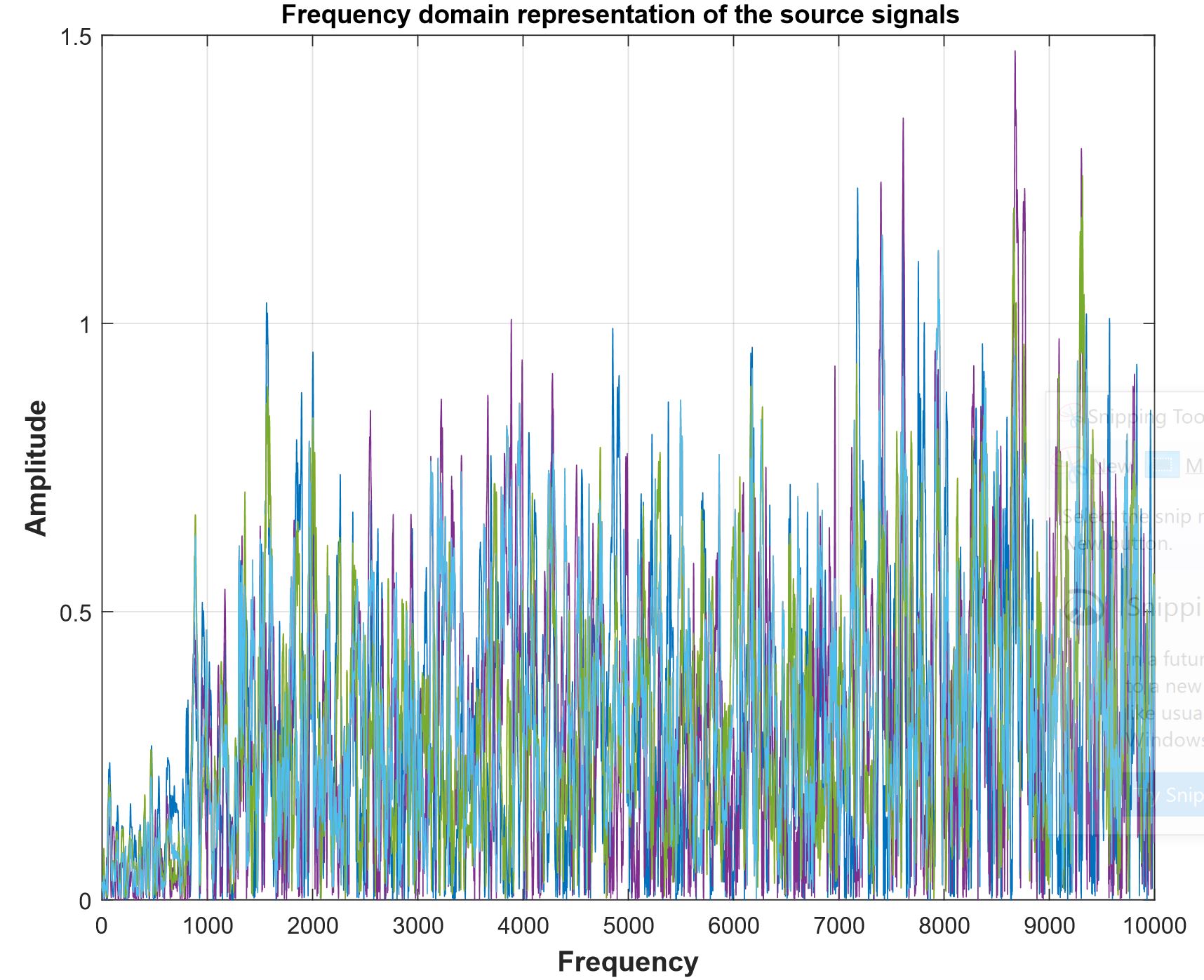}

  \caption{Frequency domain representation of the recorded audio source signals.}
    \label{Figure 4}
\end{figure}

\begin{figure}[H]
 \centering

  \includegraphics[width=5.4in]{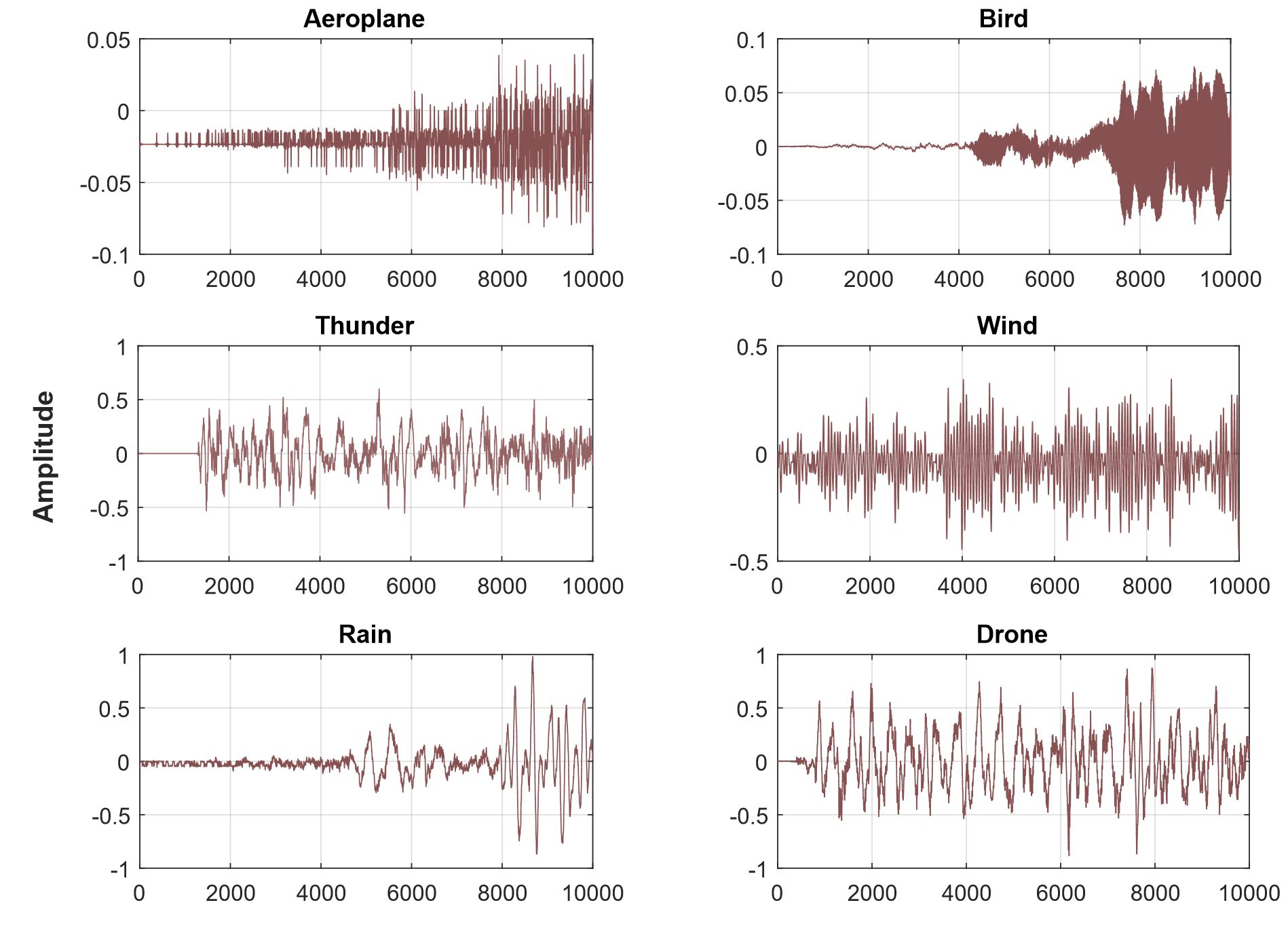}

  \caption{Audio source signals of the aeroplane, bird, wind, rain, thunder, and the AmDr recorded through microphone.}
    \label{Figure 3}
\end{figure}

\begin{figure}[H]
 \centering

  \includegraphics[width=5.4in]{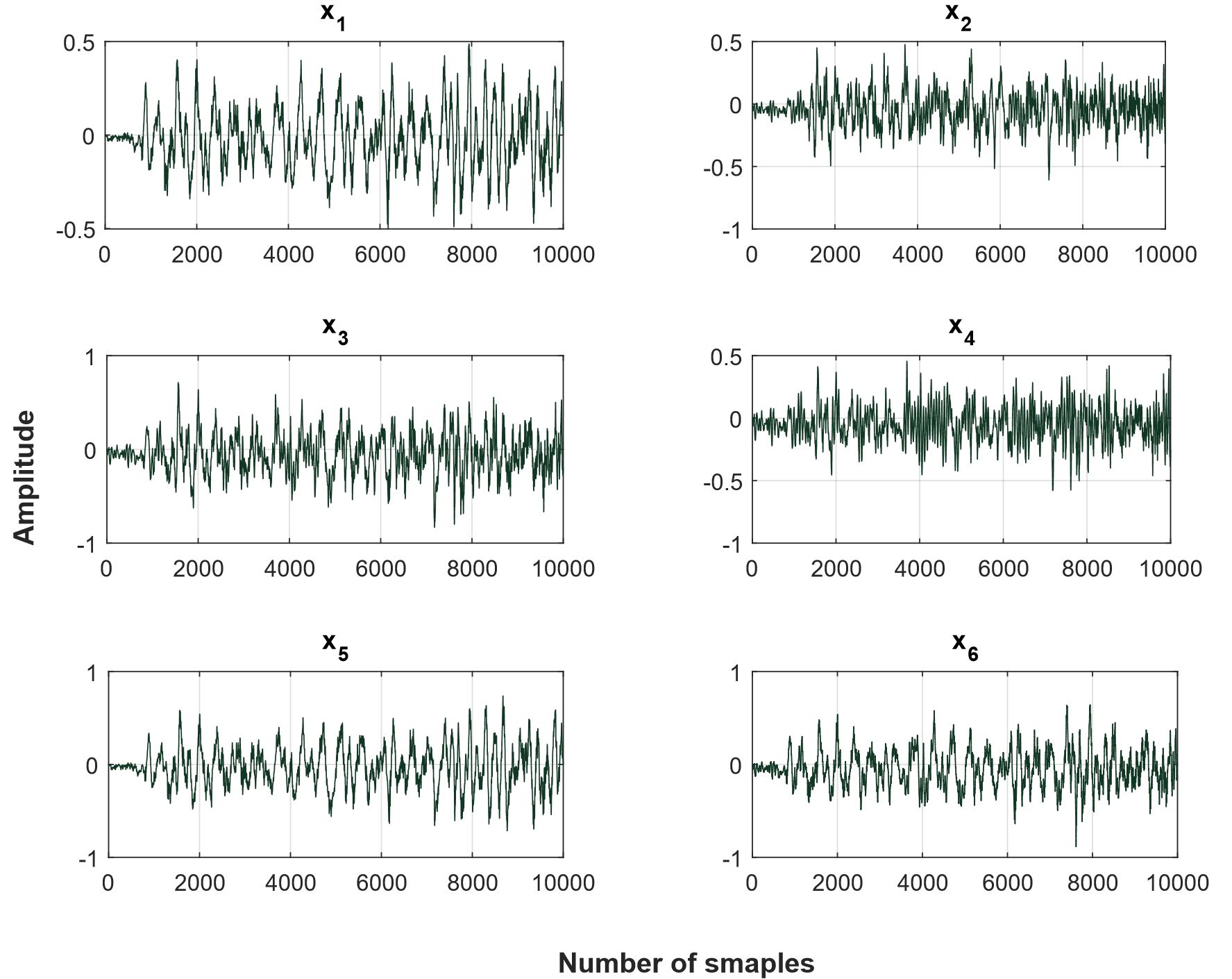}

  \caption{Mixed signals recorded through multiple microphones.}
    \label{Figure 5}
\end{figure}

In practical scenario various interfering sources exist while detecting the AmDr sound using acoustic signal processing.  Sounds like that of  aeroplanes, birds, wind, rain and thunderstorm are taken as interference sources. The independently recorded versions of these signals are shown in Figure \ref{Figure 3} in time domain and Figure \ref{Figure 4}, in frequency domain. These signals are downloaded from different databases which are freely  available for further research \cite{dbases}. From the frequency domain representation it can be observed that all the frequencies are overlap with one another and hence is difficult to separate through filtering mechanism. Further, in practical scenario the sensors will sense mixtures of all the sounds along with the AmDr sound. The recorded mixed signals from different microphones are shown in Figure \ref{Figure 5}. These signals are required to  be isolated before further processing. The ICA technique is used for separation of these mixed signals.

\section{The proposed Amateur drone detection technique} \label{sec:prop}
\begin{figure}[H]
  \centering
 
  \includegraphics[width=5.4in]{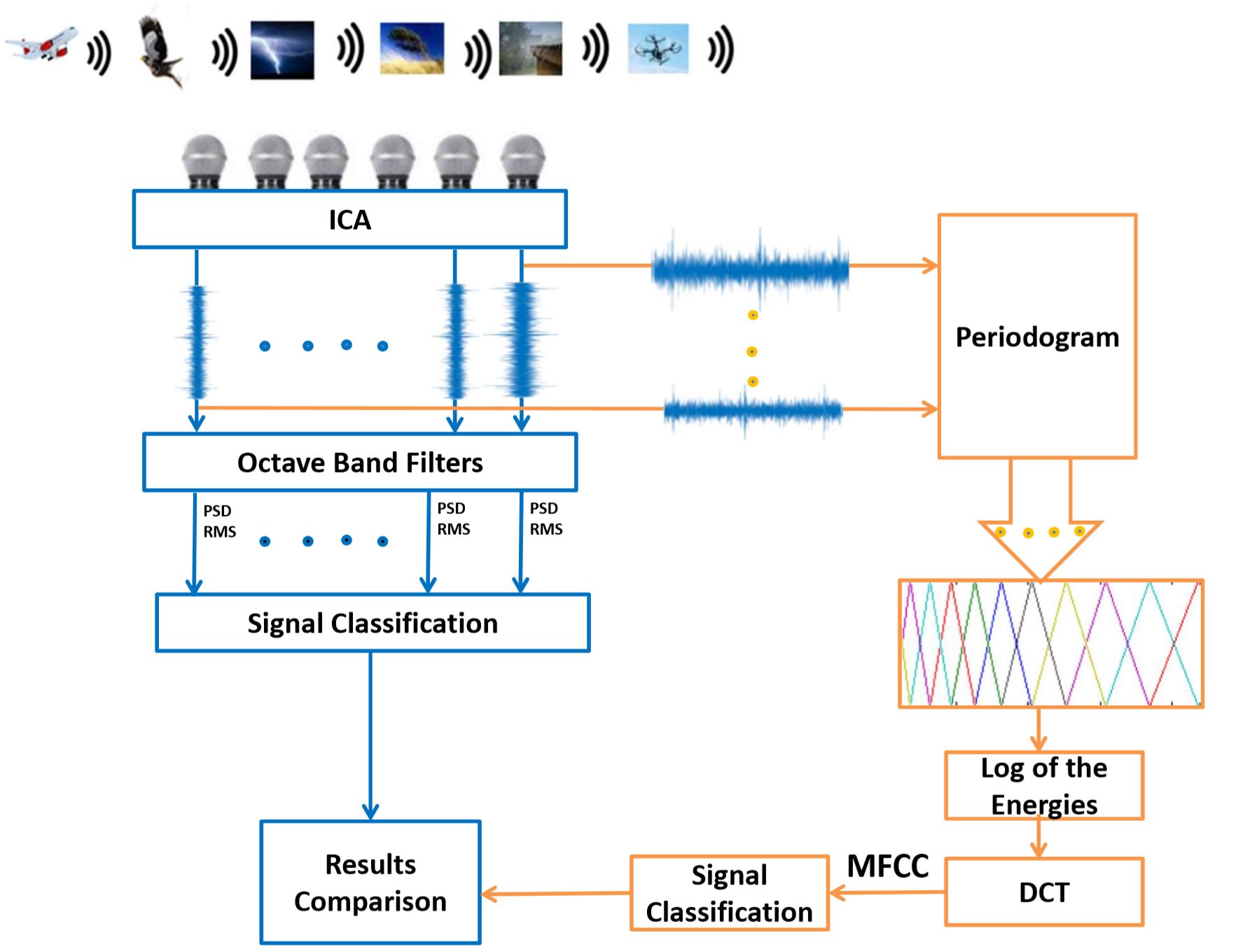}

  \caption{Block diagram representation of the proposed drone detection technique.}
    \label{Fig.proposed}
\end{figure}

The AmDr detection is a challenging task in practical scenario. The existing works considered detection of a single AmDr at a time or utilized the independently recorded signals of the AmDr and interfering sources. In practical scenario, the microphones record all the sounds present in the sensing field. Our proposed technique can detect multiple AmDrs in the presence of various interfering sources. The block diagram showing overall operation of the proposed technique is given in Figure \ref{Fig.proposed}. The mixed received signals from all the microphones are first un-mixed through using the FastICA algorithm \cite{FICA} of ICA. We select this algorithm because it is used as a benchmark in the ICA applications. The FastICA algorithm is summarized in Algorithm 1, where $E$ is the expectation operator, $g$ is the contrast function used for performance optimization of the algorithm, $P$ is the maximum number of iterations, $\textbf{W}$ is the un-mixing matrix and is inverse of the mixing matrix, and $T$ is the transpose.

 \begin{algorithm}[H]
  \vspace{\baselineskip}
\caption{The FastICA algorithm}
\label{alg1}
\begin{algorithmic}[1]
\STATE \textbf{Initialization}\\
\vspace{\baselineskip}
  \STATE Make the available data zero mean \\
  \vspace{\baselineskip}
  \STATE Whiten the data \\
  \vspace{\baselineskip}
  \STATE Choose a random initial un-mixing matrix $\textbf{W}$ of unit norm\\
  \vspace{\baselineskip}
      \FOR{$p=1$ to $P$}
      \vspace{\baselineskip}
       \STATE update the un-mixing matrix as\\
       \vspace{\baselineskip}
    $\textbf{W}_{p+1} \leftarrow E[\textbf{x}_p g(\textbf{W}^T_p \textbf{x}_p)]- E[\textbf{x}_p {g^,}(\textbf{W}^T_p \textbf{x}_p)]{\textbf{W}_p} $\\
    \vspace{\baselineskip}
\ENDFOR
\vspace{\baselineskip}
\STATE Perform un-mixing using $\textbf{Y}\leftarrow \textbf{W}\textbf{X}$
\end{algorithmic}
\end{algorithm}

The FastICA separated signals are then forwarded to octave band filtering block for calculation of the PSD.
The PSD can be calculated using Fourier Transform of the auto-correlation function as given in equation \ref{eqn:PSD}.

\begin{equation} \label{eqn:PSD}
\begin{split}
\textbf{s}_{y}(f)=\int_{-T}^{T}\textbf{r}_{y}(\tau) e^{-{j2\pi ft}}dt\\
\textbf{r}_y(\tau)=E\{\textbf{y}(t)*\textbf{y}(t-\tau)\}
\end{split}
\end{equation}
where $\textbf{s}_y(f)$ is the Fourier Transform and $\textbf{r}_y(t)$ represents the auto-correlation function  and $\textbf{y}(t)$ is the signal under test. For further details regarding the PSD the interested readers are referred to \cite{open}. In octave band, the signals are passed through different band pass filters tuned at frequencies given in Table \ref{tab:octave}. Octave bands are used for detailed analysis of the complex sounds. The audio spectrum from ~ 20 Hz to ~ 20 KHz is divided into 11 octave bands such that the highest frequency is twice the lowest frequency. However, per octave fractional bandwidth is constant and equal to 70.7\%.

\begin{table}[H]
\caption{Octave band frequencies used}
\centering 
\begin{tabular}{cccc}
\hline\noalign{\smallskip}
\textbf{Band} & \textbf{Low Freq} & \textbf{Central Freq} & \textbf{High Freq} \\ 
\noalign{\smallskip}\hline\noalign{\smallskip}
Band1 & 22.09 & 31.25 & 44.2\\ 
Band2 & 44.19 & 62.5 & 88.38\\ 
Band3 & 88.38 & 125 &  176.77\\ 
Band4 & 176.77 & 250 & 353.55\\ 
Band5 & 353.55 & 500 &	707.10\\ 
Band6 & 707.10 & 1000 & 1414.21\\ 
Band7 & 1414.21 & 2000 & 2828\\ 
Band8 & 2828.43 & 4000 & 5656\\ 
Band9 & 5656.85 & 8000 & 11313\\ 
\noalign{\smallskip}\hline
\end{tabular}
\label{tab:octave}
\end{table}

The PSD and RMS values for signal in each band are calculated and both the values are passed on to the SVM and the KNN for classification. The SVM and KNN based classification is discussed as follows. SVM was originally designed for two class classification problems ($y_j \in \{1,-1\}$). It performs classification by finding the hyper-plane that separates two classes with maximum possible margin by optimizing the following objective function using the training data

\begin{align}
&\underset{\mathbf{z}, b, \xi}{\min} \left(\frac{1}{2}\mathbf{z}^{\top}\mathbf{z} + C \sum_j \xi_j \right) \\
&s.t. ~~ y_j(\mathbf{z}\mathbf{y}_j + b) \geq 1-\xi_j, \xi_j \geq 0 \nonumber
\end{align}
Where $\mathbf{z}$ and $b$ represent the hyperplane, $C$ is the regularization constant and $\xi^j$ are used to incorporate the non-separable cases. For non-linear separation, in higher dimensions, the constraint $y_j(\mathbf{z} \phi(\mathbf{y}_j) + b) \geq 1-\xi_j, \xi_j \geq 0$ can be used. After computing the parameters of the optimal hyper-plane, the label $y_t$ of a test feature vector ${\mathbf{y}}_t$ is determined using the sign of $\frac{\mathbf{z}\mathbf{y}_t + b}{\|\mathbf{z}\|}$. Readers are referred to \cite{BookSVM,BookSVM2} for detailed explanation of SVM. In this work LibSVM \cite{LibSVM} library is used for the computation of hyper-plane parameters. Furthermore, KNN is one of the supervised learning algorithms used in data mining and machine learning, its a classifier algorithm where the learning is based on similarity index of data. KNN is based on Mahalnobis distance between two feature vectors  $\mathbf{y}_i$ and $\mathbf{y}_j$, which is defined as \cite{Mahala}.
\begin{equation}
d = (\mathbf{y}_i - \mathbf{y}_j)^{\top}\mathbf{C}^{-1}(\mathbf{y}_i - \mathbf{y}_j)
\end{equation}
where $\mathbf{C} \in\mathbb{R}^{p\times p} $ is the covariance matrix and can be calculated from the training feature vectors. Labels of training samples are assigned to the test vector ${\mathbf{y}}_t$, based on minimum distance $d$ between the training sample and the test vector ${\mathbf{y}}_t$. This procedure is then extended to a voting based $KNN$ classification.

In the second case, the MFCCs \cite{MFCC} are calculated by first finding the periodogram or PSD of  the signals. The PSD values of the signals are then passed through the Mel filter bank with frequencies as given in equation \ref{eqn:melbank}, followed by the DCT of log of the energies from Mel filer bank. Then 2 to 13 coefficients of the MFCC were used as feature vector for SVM and KNN classification.

\begin{equation} \label{eqn:melbank}
\begin{split}
H_m(k) = \begin{cases} 0, &  k<f(m-1) \\ \frac{k-f(m-1)}{f(m)-f(m-1)}, &  f(m-1)\leq k\leq f(m)\\ \frac{f(m+1)-k}{f(m+1)-f(m)}, &  f(m)\leq k\leq f(m+1)\\0, &  k>f(m+1) \end{cases}
\end{split}
\end{equation}
 Where $m$ is the number of filters $f$ is the list of $M+2$ mel spaced frequencies.

 \section{Experimental verification of the proposed work} \label{sec:Exp}
This section demonstrates the effectiveness of the proposed ICA based AmDr detection technique for audio signals. A recently published paper \cite{ad4} addressed this issue while considering the sounds of the AmDr independently of interfering sources. In practical scenario, the recorded signals are the mixtures of all the surrounding sound sources e.g., drone, birds, wind, thunderstorm, aeroplane, rain etc as shown in Figure 5. Hence, technique presented in \cite{ad4} fails in practical scenarios. It must be kept in mind that if the number of interfering sources increases then the number of recording sensors must also be increased to achieve the condition of square mixing matrix. On the other hand, frequency domain representation of the mixed recorded signals is given in Figure 7 that shows overlapped frequencies of the mixed signals. Due to frequency overlapping normal filters are un-able to separate them. We proposed an efficient technique of ICA the FastICA \cite{c3} for un-mixing of the recorded mixed sounds in the surrounding area.

\begin{figure}[H]
  \centering

  \includegraphics[width=6in]{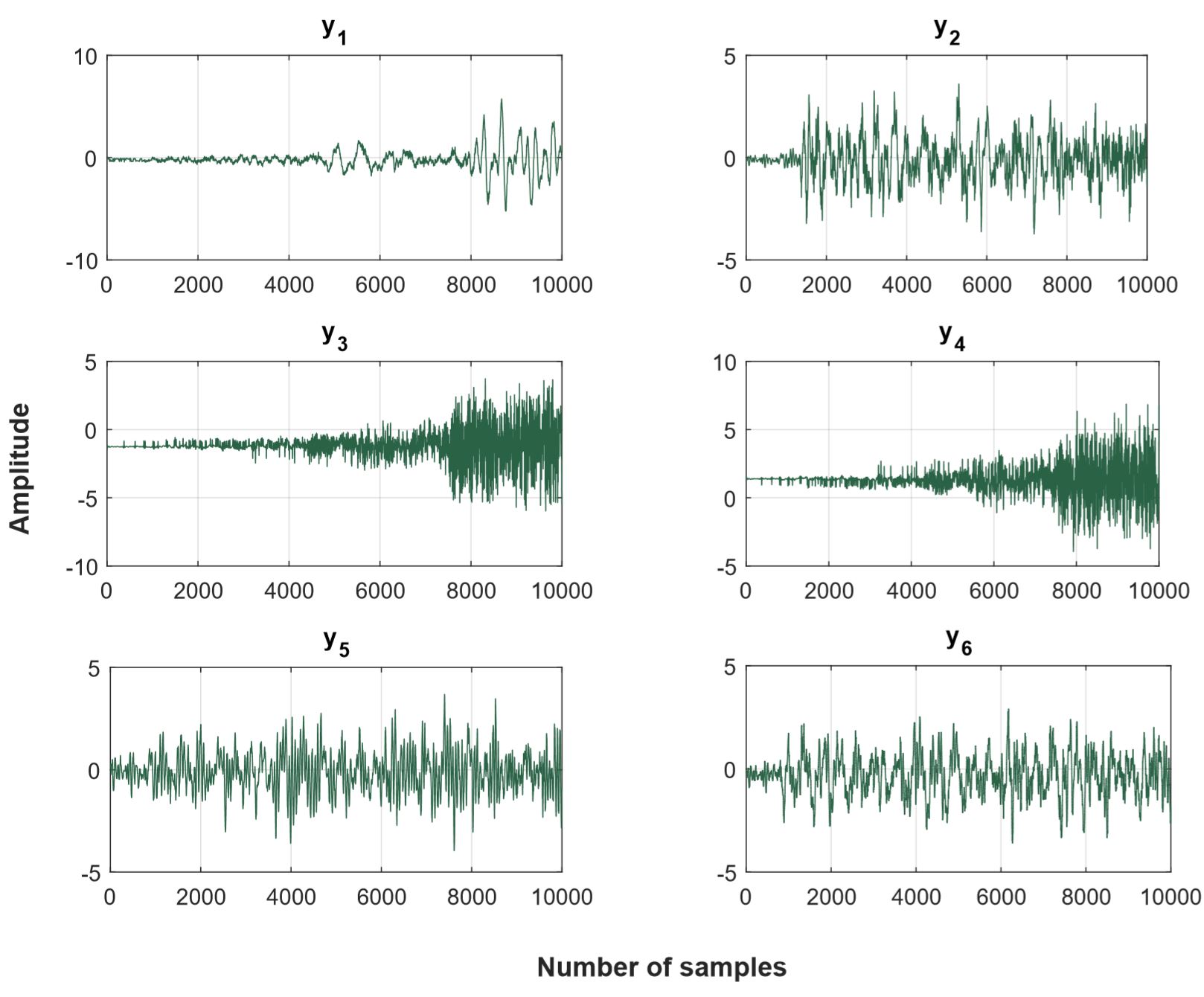}

  \caption{ICA estimated audio source signals through FastICA algorithm.}
    \label{Fig. 7}
\end{figure}

Six audio recorded mixed signals are used for simulation purpose to observe the separation performance of the FastICA algorithm. The FastICA separated signals are shown in Figure \ref{Fig. 7}.

Further, the recorded mixed data utilized for simulation contain 1000 to 10,000 samples in a single data block. Performance evaluation criterion used is signal to interference ratio (SIR). SIR is the ratio of the powers of the source signals and the estimated error signals $\textbf{s}(n)-\textbf{y}(n)$. SIR can be expressed in dB as follows

\begin{equation}
SIR_{(dB)}=\dfrac{1}{L}\sum_{n=1}^{L}\dfrac{\textbf{s}(n)^{2}}{\textbf{s}(n)-\textbf{y}(n)^{2}}
\end{equation}

\begin{figure}[H]
 \centering
  \includegraphics[width=5in]{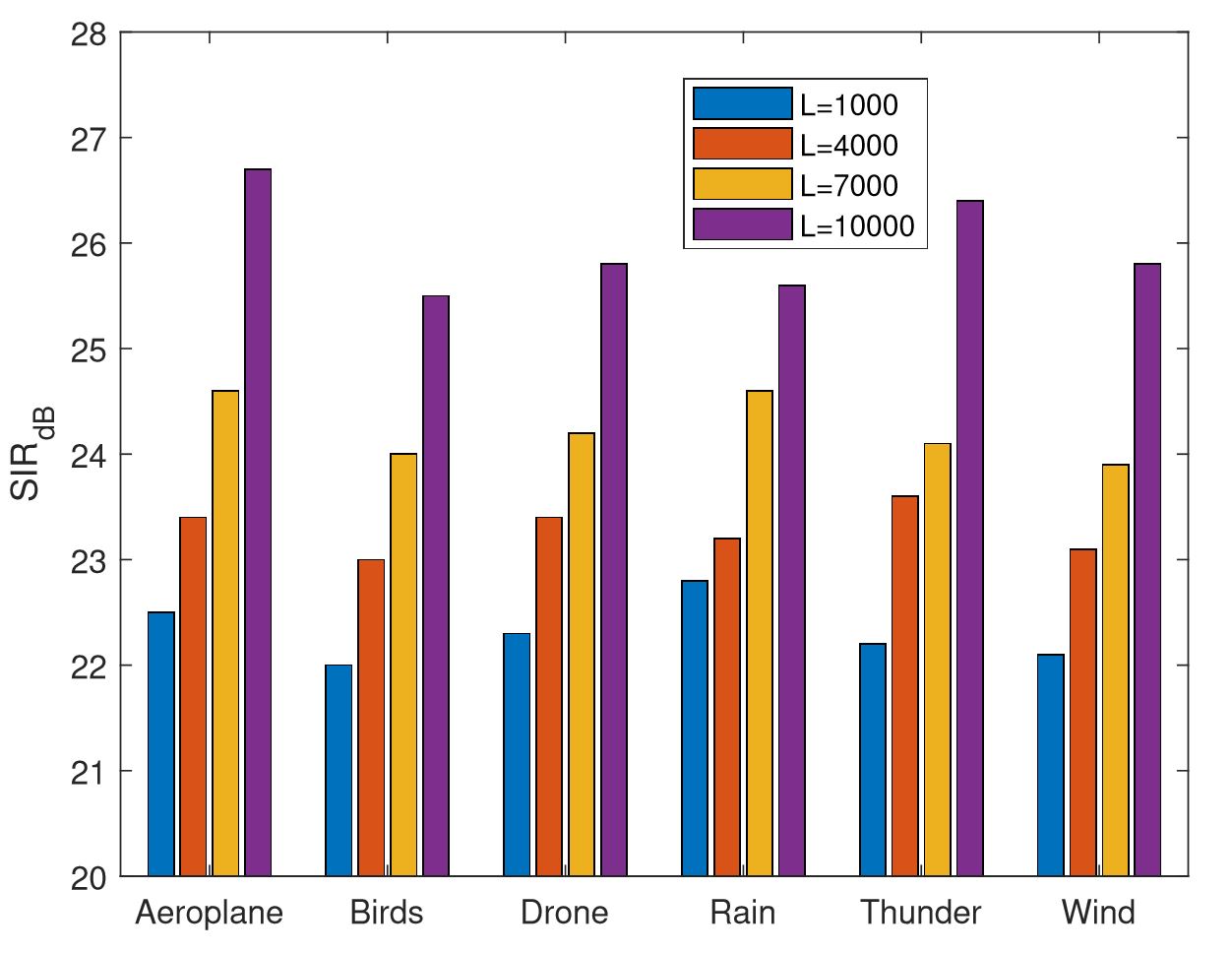}

  \caption{SIR performance of the FastICA algorithm for different data block lengths.}
    \label{Fig. SIR}
\end{figure}

The SIR performance of the FastICA algorithm is shown in Figure \ref{Fig. SIR} for all the audio signals with different data block lengths. It can be observed that the SIR performance improves with the increase in data block length from 1000 to 10000 samples.

\begin{figure}[H]
\centering
  \includegraphics[width=5.4in]{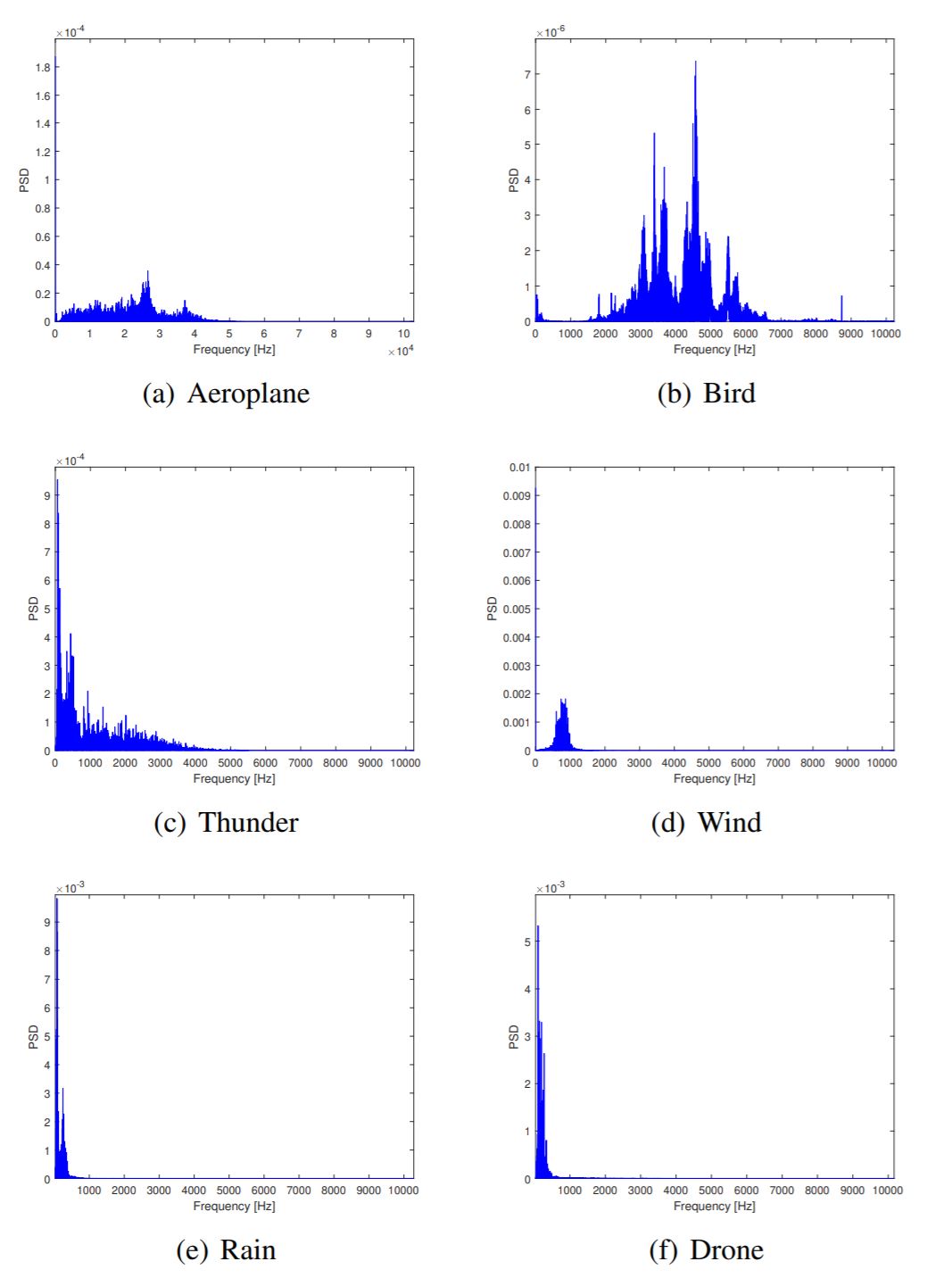}
	\caption{%
		Power Spectral Density of different audio signals }%
	\label{fig:psdall}
\end{figure}

\begin{figure}[H]
\centering
  \includegraphics[width=5.4in]{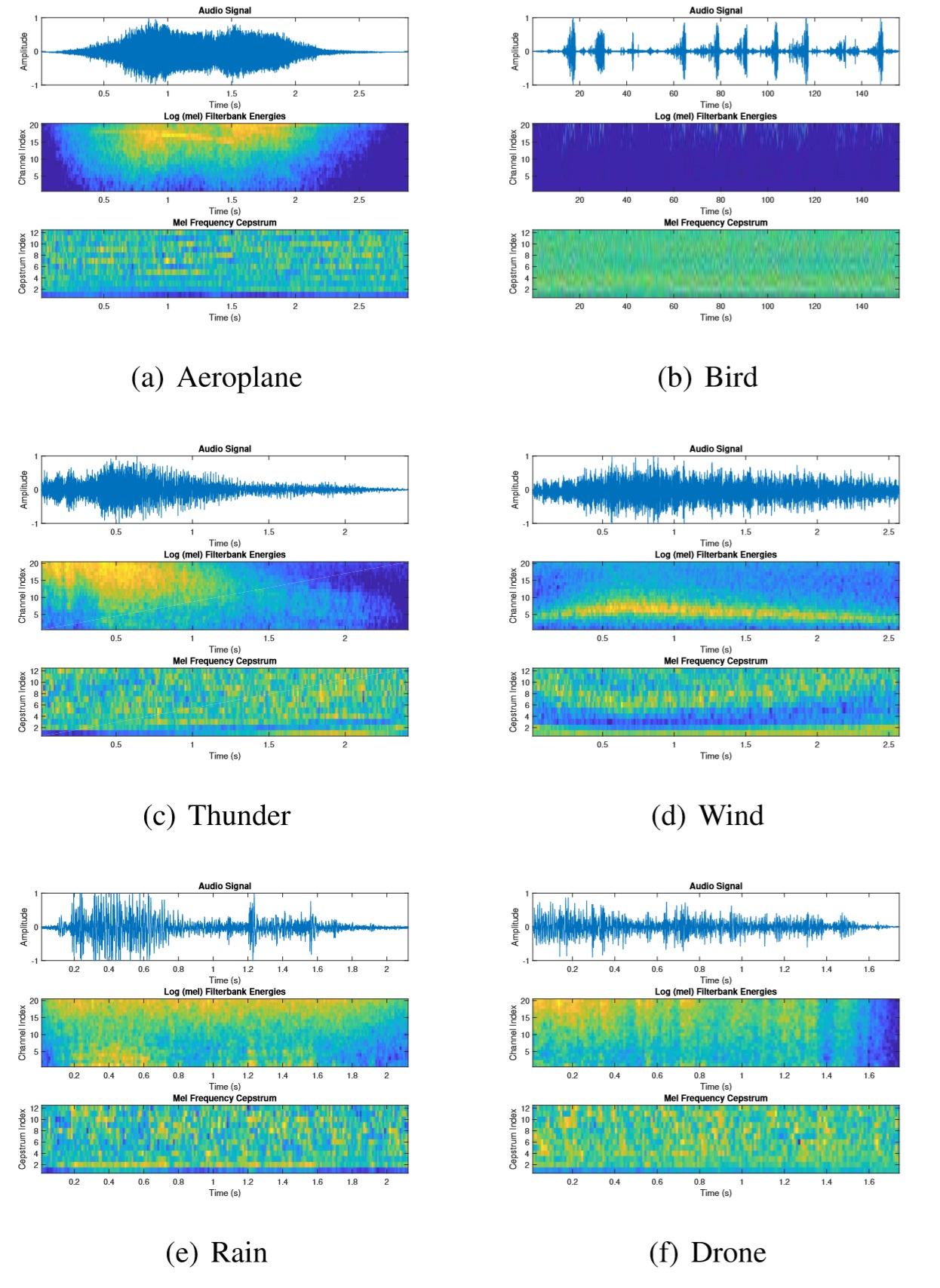}
	\caption{%
		MFCC of different audio signals }%
	\label{fig:mfccall}
\end{figure}

Classification of results after passing the mixed recorded signals through ICA algorithm and other blocks as given in Figure \ref{Fig.proposed} are tabulated in Table \ref{tab:classification}. The PSD values and the Mel energies and MFCCs are given in Figures. \ref{fig:psdall} and \ref{fig:mfccall} respectively. It can be seen from Table \ref{tab:classification} that the results obtained from the RMS values of PSD obtained gives best classification using KNN, followed by the PSD values and the Cepstral Coefficients of the Mel frequencies. Similar results are obtained while using the  SVM for classification, that is the RMS values of PSD outperforms PSD and the MFCC values of the signal under test. This table also gives the sample size used in ICA  for signal un-mixing, showing improvement in results with increase in sample size, also shown in Figure 10. This is due to the reason that increase in sample size increases the quality of unmixed signals as discussed above and hence improves the classification.

It is important to note that the computation efficiency of our proposed scheme is much less than that of MFCCs calculation as the extra operations required for DCT computation and calculating log of Mel energies are not required in case of PSD calculation from octave bands.
%
\begin{table}[H]
\caption{Classification Results}
\centering
\begin{tabular}{ccccc}
\hline\noalign{\smallskip}
 \textbf{Data Block Size}  & \textbf{Method} & \textbf{SVM} & \textbf{KNN}  \\ 
 \noalign{\smallskip}\hline\noalign{\smallskip}
\multirow{3}{*}{L=10000} & PSD & 92.57 & 97.9   \\
                  & RMS PSD & 96.1 & 99.1   \\
                  & MFCC & 88.2 & 97.4   \\ \cline{1-4}
\multirow{3}{*}{L=7000} & PSD & 91 &  97.2  \\
                  & RMS PSD & 94.9 &  98.3  \\ 
                  & MFCC & 87.6 &  97  \\ \cline{1-4}
\multirow{3}{*}{L=4000} & PSD & 90.3 &  96.7  \\ 
                  & RMS PSD & 94.1 & 98   \\
                  & MFCC & 87.0 & 96.7   \\ \cline{1-4}
\multirow{3}{*}{L=1000} & PSD & 89.7 & 96.0   \\ 
                  & RMS PSD & 93.3 & 97.1   \\ 
                  & MFCC & 86.8 &  95.3  \\ 
\noalign{\smallskip}\hline
\end{tabular}
\label{tab:classification}
\end{table}

in this work, we developed a practically applicable technique of the AmDr detection that can detect multiple AmDrs in the field. It is also important to mention that we are the first to introduce the concept of blind separation in the field of AmDr detection. In the proposed design it is important to separate the mixed recorded signals with good quality results. As it is known that performance of the ICA algorithms degrade while reducing lengths of the processing data blocks i.e., sufficient number of samples must exist in the processing data blocks. In case of smaller data blocks performance of the classification algorithms also degrade due to low quality separated signals.
\section{Conclusion} \label{sec:conclude}
Unmanned Air Vehicles are constantly being used by public for different purposes, like picturizing different events to delivering light weight objects. these vehicles can also be used for spying and other threatening activities, If not properly observed. Therefore, its timely detection is one of the important task for public safety and security. In this research work we presented a new approach for sound based detection of AmDrs. The audio signals from  different sources in the area under observation were first unmixed, then passed through  filters separating the signal into different octave bands. The PSD and RMS values of PSD were calculated. The PSD and RMS values of PSD were then fed to different machine learning algorithms for classification into to drone and non-drone sounds. The results were then compared with the MFCC based AmDr detection techniques available in literature, however, it is important to note that here the MFFCs of the unmixed signals were used, which is a departure from the MFFCs algorithm utilization in drone detection, found in literature. The simulation results show that our proposed algorithm outperformed the MFCC based classification algorithm and showed 99.1\% accuracy. It is also  found that the accuracy of drone detection increases with increase in the sample size for ICA. The proposed algorithm can be easily be adopted in real life scenarios as the octave band filtering and PSD calculation require much less computation as compared to other techniques like image based detection, video based detection and even MFCC based adio detection.



  \bibliographystyle{elsarticle-num}

\begin{thebibliography}{}
\bibitem{wc1}
A. Al-Hourani, S. Kandeepan, and A. Jamalipour, Stochastic geometry
study on device-to-device communication as a disaster relief solution,
IEEE Trans. Veh. Technol., vol. 65, no. 5, pp. 3005-3017, 2016.

\bibitem{wc2}
Z. Kaleem and K. Chang, "Public safety priority-based user association
for load balancing and interference reduction in PS-LTE systems", IEEE
Access, vol. 4, pp. 9775-9785, 2016.
\bibitem{wc3}
Z. M. Fadlullah, D. Takaishi, H. Nishiyama, N. Kato, and R. Miura, "A
dynamic trajectory control algorithm for improving the communication
throughput and delay in UAV-aided networks", IEEE Network, vol. 30,
no. 1, pp. 100-105, 2016.
\bibitem{wc4}
A.-H. Akram, S. Chandrasekharan, G. Kaandorp, W. Glenn, A. Jamalipour,
and S. Kandeepan, "Coverage and rate analysis of aerial base
stations", IEEE Trans. Aerosp. Electron. Syst., vol. 52, no. 6, pp. 3077-3081, 2016.
\bibitem{wc5}
D. He, S. Chan, and M. Guizani, "Drone-assisted public safety networks:
The security aspect", IEEE Commun. Mag., vol. 55, no. 8, pp. 218-223,
2017.
\bibitem{wc6}
Haubeck, K., and T. Prinz. "A UAV-based low-cost stereo camera system for archaeological surveys, Experiences From Doliche (Turkey)", Int. Arch. Photogramm. Remote Sens. Spat. Inf. Sci (2013): 195-200.
\bibitem{wc7}
Eltner, A., C. Mulsow, and H. G. Maas. "Quantitative measurement of soil erosion from TLS and UAV data", ISPRS-International Archives of the Photogrammetry, Remote Sensing and Spatial Information Sciences 1.2 (2013): 119-124.
\bibitem{wc8}
D. Brucas, et al. "Implementation and Testing of Low Cost UAV Platform for Orthophoto Imaging", International Archives of the Photogrammetry, Remote Sensing and Spatial Information Sciences XL-1 W 2 (2013): 55-59.
\bibitem{wc9}
Rokhmana, Catur Aries, "The potential of UAV-based remote sensing for supporting precision agriculture in Indonesia", Procedia Environmental Sciences 24 (2015): 245-253.
\bibitem{wc10}
Ballarin, M., C. Balletti, and F. Guerra. "Action cameras and low-cost aerial vehicles in archaeology", Videometrics, Range Imaging, and Applications XIII. Vol. 9528. International Society for Optics and Photonics, 2015.
\bibitem{wc11}
Longhi, M., et al. "Rfidrone: Preliminary experiments and electromagnetic models." (EMTS), IEEE International Symposium on Electromagnetic Theory. , 2016.
\bibitem{wc12}
Huang, Yo-Ping, Lucky Sithole, and Tsu-Tian Lee, "Structure From Motion Technique for Scene Detection Using Autonomous Drone Navigation", IEEE Transactions on Systems, Man, and Cybernetics: (2017).
\bibitem{wc13}
Ma Lei, et al. "Using unmanned aerial vehicle for remote sensing application", IEEE 21st International Conference on Geoinformatics, 2013.
\bibitem{wc14}
McGarey, Patrick, and Srikanth Saripalli. "Autokite experimental use of a low cost autonomous kite plane for aerial photography and reconnaissance", IEEE International Conference on Unmanned Aircraft Systems (ICUAS), 2013.
\bibitem{wc15}
V. Baiocchi, D. Dominici, and M. Mormile, "UAV application in post-seismic environment", Int. Arch. Photogramm. Remote Sens. Spatial Inf. Sci., XL-1 W 2 (2013): 21-25.
\bibitem{wc16}
Burdziakowski, Pawel. "Low cost hexacopter autonomous platform for testing and developing photogrammetry technologies and intelligent navigation systems." (2017): 1-6.
\bibitem{wc17}
Gini, Rossana, et al., "Use of unmanned aerial systems for multispectral survey and tree classification: A test in a park area of northern Italy", European Journal of Remote Sensing, 47.1 (2014): 251-269.

\bibitem{wc18}
Gynnild, Astrid, "The Robot Eye Witness: Extending visual journalism through drone surveillance", Digital journalism, 2.3 (2014): 334-343.
\bibitem{com1}
Aggarwal, Shubhani, and Neeraj Kumar, "Path planning techniques for unmanned aerial vehicles: A review, solutions, and challenges", Computer Communications (2019).
\bibitem{com2}
Saharan, Sandeep, Seema Bawa, and Neeraj Kumar, "Dynamic pricing techniques for Intelligent Transportation System in smart cities: A systematic review", Computer Communications (2019).

\bibitem{ad1}
Z. Kaleem, and M. H. Rehmani, "Amateur Drone Monitoring: State-of-the-Art Architectures, Key Enabling Technologies, and Future Research Directions", IEEE Wireless Communications 25.2 (2018): 150-159.
\bibitem{ad2}
Azari, Mohammad Mahdi, et al. "Key technologies and system trade-offs for detection and localization of amateur drones", IEEE Communications Magazine, 56.1 (2018): 51-57.
\bibitem{ad3}
Ding, Guoru, et al. "An amateur drone surveillance system based on the cognitive Internet of Things", IEEE Communications Magazine, 56.1 (2018): 29-35.
\bibitem{ad5}
G. Ding, et al., "An amateur drone surveillance system based on the cognitive internet of
things", IEEE Commun. Mag., vol. 56, no. 1, pp. 29-35, 2018.
\bibitem{ad6}
H. Liu, Z. Wei, Y. Chen, J. Pan, L. Lin, and Y. Ren, Drone detection
based on an audio-assisted camera array", IEEE 3rd International Conference
on Multimedia Big Data (BigMM), 2017, pp. 402-406.
\bibitem{ad4}
M. Z. Anwar, Z. Kaleem, "Machine Learning Inspired Sound-based Amateur Drone Detection for Public Safety Applications", DOI: 10.1109/TVT.2019.2893615,  December 2018.
\bibitem{ad10}
J. Kim, C. et. al., "Realtime UAV sound detection and analysis system", IEEE Sensors Applications
Symposium (SAS), 2017, pp. 1-5.
\bibitem{ad7}
S. J. Lee, J. H. Jung, and B. Park, "Possibility verification of drone detection
radar based on pseudo random binary sequence", IEEE International
Conference on SoC Design(ISOCC), 2016, pp. 291-292.
\bibitem{ad8}
J. Drozdowicz, M. Wielgo, P. Samczynski, K. Kulpa, J. Krzonkalla,
M. Mordzonek, M. Bryl, and Z. Jakielaszek, "35 GHz FMCW drone
detection system", IEEE 17th International Radar Symposium (IRS). ,
2016, pp. 1-4.
\bibitem{ad9}
Ryden, Henrik, Sakib Bin Redhwan, and Xingqin Lin, "Rogue drone detection: A machine learning approach", IEEE Wireless Communications and Networking Conference (WCNC), 2019.
\bibitem{ad11}
J. Mezei and A. Molnar, "Drone sound detection by correlation", IEEE 11th International Symposium on Applied Computational Intelligence and Informatics (SACI), 2016, pp. 509-518.
\bibitem{ad12}
T. Muller, "Robust drone detection for day/night counter-UAV with static VIS and SWIR cameras," Proc. SPIE 10190, Ground/Air Multisensor Interoperability, Integration, and Networking for Persistent ISR VIII, 2017.
\bibitem{ad13}
L. Shi, I. Ahmad, Y. He, and K. Chang, "Hidden markov model based drone sound recognition using MFCC technique in practical noisy environments", Journal of Communications and Networks, vol. 20, no. 5, pp. 509-518, 2018.
\bibitem{xshi}
Shi, Xiufang, et al. "Anti-drone system with multiple surveillance technologies: Architecture, implementation, and challenges." IEEE Communications Magazine 56.4 (2018): 68-74.
\bibitem{c2}
Formaggio, Emanuela, et al. "Time frequency modulation of ERD and EEG coherence in robot-assisted hand performance", Brain topography, 28.2 (2015): 352-363.
\bibitem{c3}
Z. Uddin, et al. "Independent component analysis based MIMO transceiver with improved performance in time varying wireless channels", KSII Transactions on Internet and Information Systems (TIIS), 9.7, (2015): 2435-2453.
\bibitem{c4}
Z. Uddin, A. Ahmad, and M. Iqbal. "ICA Based MIMO Transceiver For Time Varying Wireless Channels Utilizing Smaller Data Blocks Lengths." Wireless Personal Communications, 94.4, (2017): 3147-3161.
\bibitem{c5}
Z. Uddin, et al., "Modified Infomax algorithm for smaller data block lengths," Springer Wireless Personal Communications, vol. 87, no. 1, pp- 245-267, March 2016.
\bibitem{c6}
Z. Uddin, et al., "Applications of Independent Component Analysis in Wireless Communication Systems", Wireless Personal Communication , vol. 83, Issue 4, pp: 2711-2737, 2015
\bibitem{LMCC}
A. Kumar, S. S. Rout, and V. Goel, "Speech Mel frequency cepstral coefficient feature classification using multi level support vector machine", 4th IEEE Uttar Pradesh Section International Conference on Electrical, Computer and Electronics (UPCON), 2017, pp. 134-138.
\bibitem{LSVM}
L. Grama, L. Tuns, and C. Rusu, "On the optimization of SVM kernel parameters for improving audio classification accuracy", 14th International Conference on Engineering of Modern Electric Systems
(EMES), June 2017, pp. 224-227
\bibitem{d}
Norman-Haignere, Sam, Nancy G. Kanwisher, and Josh H. McDermott, "Distinct cortical pathways for music and speech revealed by hypothesis-free voxel decomposition", Neuron, 88.6, (2015): 1281-1296.
\bibitem{e}
Rimmele, Johanna M., et al., "The effects of selective attention and speech acoustics on neural speech-tracking in a multi-talker scene", Cortex, 68, (2015): 144-154.
\bibitem{dbases}
https://www.soundsnap.com/tags, https://freesound.org/browse/tags
\bibitem{FICA}
Uddin, Zahoor, et al. "Adaptive Step Size Gradient Ascent ICA Algorithm for Wireless MIMO Systems", Mobile Information Systems, vol. 2018 (2018).
DOI: https://doi.org/10.1155/2018/7038531
\bibitem{open}
Oppenheim, A.V.,and Schafer, R.W., "Discrete-Time Signal Processing", 3rd edn. Prentice Hall Press, Upper Saddle River, NJ, USA (2009)
\bibitem{BookSVM2}
Scholkopf, Bernhard and Smola, Alexander J., "Learning with Kernels: Support Vector Machines, Regularization, Optimization, and Beyond." MIT (2001)
\bibitem{Mahala}
Theodoridis, S., Koutroumbas, K., "Pattern Recognition", Academic, Boston, USA (2010)
\bibitem{MFCC}
Steven .B.Davis and paul Mermelstein, "Comparison of parametric representations for monosyllabic word recognition in continuously spoken sentences", IEEE Trans. on Acoust Speech Signal Processing, 28.4, (1980): 357-366
\bibitem{BookSVM}
Cristianini, Nello and Shawe Taylor, John, "An Introduction to Support Vector Machines and other kernel-based learning methods", Cambridge University (2003)
\bibitem{LibSVM}
Chang, Chung and Lin, Jen, "LIBSVM: A Library for Support Vector Machines."  ACM Trans. on Intell. Syst. Tech. 3.2 (2011): 1-27

\end{thebibliography}


\end{document}